Vertical modes of two-dimensional dust coulomb clusters in complex plasmas

K. Qiao and T. W. Hyde

Center for sstrophysics, space physics and engineering research, baylor
university, waco, TX, 76798-7310, USA

The vertical as well as horizontal oscillation modes in a thermally excited two-dimensional (2D) dust coulomb cluster were investigated for particle numbers between $N = 3$ and $N = 150$ using both a box_tree simulation and an analytical method. The horizontal mode spectra is shown to agree with published results while the vertical mode spectra obtained from the box_tree simulation and the analytical method are shown to agree with one another. The maximum frequency of the vertical modes is shown to be the vertical oscillation frequency of the whole system acting as a solid plane $\omega_{z0}$ and the frequency is shown to decrease as the mode number $l$ decreases. The oscillation patterns of the vertical modes are also investigated. For clusters with large numbers of particles, the high frequency modes are shown to have oscillation patterns similar in shape to Bessel-Fourier functions with various index $m$ and $n$. For specified values of $n$, modes with higher $m$ (except for $m = 0$) have lower frequencies and for specified $m$, modes with higher $n$ have lower frequencies. For low frequency modes, the largest amplitude particle motion is concentrated in a few inner rings with the outer rings remaining almost motionless. This is in contrast to the horizontal modes where the strongest motion of the particles is concentrated in the inner rings at high frequency. For clusters with small numbers of particles ($N \leq 20$), these modes where the strongest vertical motion is concentrated in the inner rings does not exist resulting in oscillation magnitudes for all the modes appearing in the shape of Bessel-Fourier functions.

## I. Introduction

Plasma crystals can form in complex plasmas consisting of micron sized dust particles immersed in an ion-electron plasma.[1-3] In a complex plasma, dust particles generally become negatively charged[4] with the interaction potential between any two particles a Yukawa potential of the form

$$v(r) = q\exp(-r/\lambda_D)/4\pi\varepsilon_0 r, \qquad (1)$$

where $q$ is the dust particle charge, $r$ is the distance between any two particles and $\lambda_D$ is the dust Debye length. In the typical experimental environment on earth, a weakly ionized plasma is created via rf discharge and dust particles introduced to this plasma are levitated in the sheath region above the lower electrode due to dc self-bias.[1-3] Once levitated, these dust particles are confined in the vertical direction within a potential well created primarily by the sum of the gravitational and electrostatic forces in the sheath[5,6] while they are constrained in the horizontal direction by a parabolic electrostatic potential formed either by a containing ring placed on the lower electrode or by shaping the electrode itself.[7,8] Dust particles acted upon in such a manner form ordered lattice structures often called plasma crystals.[9]

Plasma crystals such as those discussed above form from large numbers of particles (>10,000) and thus, can usually be considered as infinite lattices for modeling purposes. However, when there are only a small number (tens) of particles introduced to the complex plasma system, instead of forming plasma crystals they often form much smaller (and much different) ordered structures. These resulting structures are called dust Coulomb clusters and must be treated as finite systems.

As mentioned above, particles introduced to the complex plasma system are horizontally confined in a shallow parabolic potential well.[10,11] The potential well in the vertical direction has been shown experimentally to be parabolic in nature,[5,12] and is generally much stronger than that causing the horizontal confinement. Thus, the total external potential energy can be modeled as

$$E_{ext}(x,y,z) = \frac{m}{2}\left[\omega_{xy0}^2(x^2+y^2)+\omega_{z0}^2 z^2\right]. \quad (2)$$

where $z$ is the particle height and $\omega_{xy0}^2$ and $\omega_{z0}^2$ are the measure of the strength of the horizontal and vertical confinements respectively. Combining the external potential energy with the interparticle Yukawa potential energy, the total energy of the system can now be represented as

$$E(x,y,z) = \frac{m}{2}\sum_{i=1}^{N}\left[\omega_{xy0}^2(x_i^2+y_i^2)+\omega_{z0}^2 z_i^2\right]+\frac{Z^2 e^2}{4\pi\varepsilon_0}\sum_{i>j}^{N}\frac{1}{r_{ij}}\exp\left(-\frac{r_{ij}}{\lambda_D}\right). \quad (3)$$

The normal modes for clusters formed in this manner can then be calculated employing the dynamical matrix

$$E_{\alpha\beta,ij} = \frac{\partial^2 E}{\partial r_{\alpha,i} \partial r_{\beta,j}}, \tag{4}$$

where α and $\beta = x, y$ and $i, j$ denote particle number. Once established, the eigenvalues of this matrix represent the normal mode frequencies (ω) of the 3N modes with the matrix eigenvectors describing the mode oscillation patterns.[11]

Recently, there has been an increased interest in the overall structure,[13] phase transition mechanisms,[14] and mode spectra[11,15] for 2D dust coulomb clusters. (Clusters most easily form 2D systems due to the strong vertical confinement mentioned above and the fact that they form from small numbers of particles.) The horizantal modes have recently been investigated intensely[11,15]. However, the total mode spectra for 2D dusters should consist of modes not only involving horizontal particle motion but vertical particle motion as well. Although dust lattice wave modes created by such vertical particle motion has already been investigated for both one-dimensional (1D)[12] and 2D[17,18] plasma crystals, it has not to our knowledge yet been examined for dust coulomb clusters accordingly. Accordingly in this research, both the vertical and horizontal modes are obtained for particle numbers between $N = 3$ and $N = 150$ employing a box_tree simulation[17-21] of thermally excited 2D dust coulomb clusters. The horizontal mode spectra obtained is then compared with previously published experimental and theoretical results[11] while the vertical mode spectra is analyzed and compared with corresponding analytical results.

II. Numerical simulation and results

The formation of dust coulomb clusters for all system particle numbers falling between $N = 3$ and $N = 150$ were simulated using the box_tree code.[17-21] For this simulation, dust particles are assumed to have constant and equal charges of $q = 3.84 \times 10^{-15} C$, equal masses of $m_d = 1.74 \times 10^{-12} kg$ and equal radii of $r_0 = 6.5 \mu m$. The dust Debye length is $\lambda_D = 0.57 mm$. The interparticle potential is modeled as a Yukawa potential of the form given by Eq. (1) while the external potential within the sheath is assumed to be parabolic in nature as given by Eq. (2) with $\omega_{xy0}^2 = 2.21$ and $\omega_{z0}^2 = 221.07$. Thermal equilibrium (~at approximately room temperature) was established by introducing collisions with room temperature neutral gas particles.

Under these conditions, an initially random distribution of dust particles formed ordered dust coulomb clusters as shown in Fig. 1 (a-c). Once established, the thermal motion of the dust particles around their equilibrium positions was tracked for 32 seconds with output data files created every 0.04 second yielding a total of 800 data files. These data were used to obtain the normal mode spectra employing the method given by Melzer,[11] taking into account not only the horizontal but also the vertical motion of all cluster particles. Additionally, the normal mode spectra were also calculated employing Eqs. (3) and (4) using the equilibrium positions of the particles. The spectra of the normal modes, including both the horizontal and vertical modes are given in Fig. 1 (d-f). As can be seen, the simulation results shown in the intensity graphs are in excellent agreement with the theoretical results represented by the solid dots. The spectra for the vertical normal modes are shown in Fig. 1 (g-i). The maximum frequency for a given vertical mode is known as the vertical oscillation frequency. For this case it was found to be $\omega_{z0} = 14.87 s^{-1}$, which corresponds to the oscillation frequency for a vertical oscillation of the entire particle system acting as a solid plane. As can be seen, the mode frequency decreases as the mode number *l* decreases.

It is interesting to note that for small numbers of particles, i.e. $N < 47$, the horizontal and vertical spectra can appear as two separate branches with the horizontal mode spectra in agreement with previous results.[11] As the overall particle number increases ($N \geq 47$), the minimum frequency for the vertical mode decreases until it falls below the maximum frequency for the horizontal modes, and a merging of the two branches occurs. In either case when considered separately, the horizontal mode spectra remains in agreement with previously published results[11] and the frequency of the vertical modes still decreases as the mode number decreases, with the maximum frequency remaining the vertical oscillation frequency, $\omega_{z0}$.

An investigation of the vertical mode oscillation patterns can be conducted by examining the eigenvectors $\vec{e}_{i,l}$ of the dynamical matrix generated by Eq. (4), and determining the oscillation amplitude and direction for particle *i* and mode number *l*. A detailed examination of this sort for the vertical modes obtained for the dust clusters mentioned above (particle number $3 < N < 150$) has been conducted. As mentioned previously, the highest frequency mode is the mode corresponding to a vertical oscillation of the entire system of particles as a solid plane. This frequency is always equal to $\omega_{z0}$ and is independent of particle number. The modes with the second and third highest frequencies correspond to rotational oscillations of the system (again as a solid plane) around two different horizontal axes. For a continuous (isotropic) system, these two modes are degenerate; in the system modeled above, they have close but slightly different frequencies due to the anisotropic nature of the cluster. For example, when $N = 150$, they have frequencies of 14.7935s$^{-1}$ and 14.7934s$^{-1}$ respectively. These three modes were found to exist for all clusters examined and are shown in Fig. 2 for the values of $N = 3$ and $N = 150$.

As the frequency and mode number $l$ decrease, the vertical modes begin to exhibit more complex patterns of oscillation. This can be seen in Figures 3 and 4 which shows the oscillation patterns for the highest frequency vertical modes after the three modes described above ($N = 150$). Interestingly, as can be seen in Figure 3 and 4, oscillation pattern for a specific mode assumes the form of a Bessel function $J_m(k_{mn}r)$ in the radial direction $r$ and a Fourier function $e^{im\theta}$ in the angular direction $\theta$. (In this case, $m$ is equal to the number of periods in the angular direction while $n$ is determined by the boundary conditions on the radial direction.)

As shown in the figure, boundary conditions at the cluster's edge are not well represented by closed boundary conditions, for which the magnitude of the oscillation should be zero, or by free boundary conditions, for which the slope would be zero. Instead, the slope at the boundary generally assumes the form of either a maximum or a minimum, i.e., the third derivative of the particle displacement is zero. Thus, if the first zero point of the third derivative of the particle displacement is at the boundary, $n = 1$; if the second zero point is at the boundary, $n = 2$ … and so on. In Figures 3 and 4, it can also be seen that for a specified value of $n$, modes with higher $m$ values (excluding the case where $m = 0$) have lower frequencies while for a specified value of $m$, modes with higher $n$ will have lower frequencies. Similar to the second and third highest frequency modes described above, for each mode with $m \neq 0$, there is a second mode having the same $m$ and $n$. As before, their frequencies are close but slightly different due to the anisotropic nature of the cluster. It can also be seen in Fig. 3 (p) that when *m* is greater than some maximum (in this case, $m \geq 12$), even for $n = 1$, the system becomes almost flat due to the finite size of the cluster and the pattern becomes hard to discern. For the same reason, as frequencies decrease to values lower than $13.5 s^{-1}$, there are multiple peaks and valleys within a relatively small area and the pattern again becomes difficult to recognize.

At the other limit, as shown in Figure 5, for the lowest frequencies (N = 150) peaks and valleys only appear at the center of the cluster while all other areas of the cluster remain flat. Thus, the maximum energy (amplitude) for vertical motion of the particles remains concentrated across a few inner rings with the outer rings remaining almost motionless in the vertical direction. This is contrary to the manner in which horizontal modes act where it is for the highest frequency modes that the particle motion is concentrated within the inner two rings and the outer rings remain more or less motionless.[11] Taken together, for a 2D dust cluster with large particle numbers, horizontal particle motion is distributed throughout the cluster while vertical motion remains concentrated at the cluster's center for low frequencies with exactly the opposite occurring for high frequencies.

For clusters with small numbers of particles, i.e., $N \leq 20$, the oscillation magnitude for all modes appears in the shape of the Bessel-Fourier functions. Table 1 shows the *m*, *n* indices for all vertical modes as a function of the increasing mode number *l* (decreasing frequency). As can be seen, the mode order is almost the same as for large clusters ($N = 150$); however, due to the finite structure of the cluster, this order breaks down for specific cases. For example, for clusters with large numbers of particles ($N = 150$), the $m = 0$, $n = 1$ mode falls after the two $m = 3$, $n = 1$ modes and before the two $m = 4$, $n = 1$ modes. However, it falls before the two $m = 3$, $n = 1$ modes for $N = 8, 20$, between the two $m = 3$, $n = 1$ modes for $N = 17, 19$ and after the two $m = 4$, $n = 1$ modes for $N = 11, 12, 14$.

Of the modes for $N \leq 20$, there are three modes which can not be explained by a Bessel-Fourier function: the final mode for $N = 14$ as shown in Fig.6 (a) and the two consecutive modes between the $m = 4$, $n = 1$ mode and the $m = 6$, $n = 1$ mode for $N = 19$, as shown in Fig.6 (b), (c). The first of these may actually be the $m = 2$, $n = 2$ mode simply deformed due to the finite lattice structure. The latter two modes appear as Fourier functions but only in one direction and thus are not solutions for a 2D circular membrane. This is to be expected since the $N = 19$ cluster has a stable hexagonal rather than circular structure.

Finally as shown in Fig.7, for the $N = 21$ cluster the vertical mode with the lowest frequency has a peak-valley pair located at the center while all other regimes are vertically motionless. Again, as the particle number continue to increase ($N > 21$), additional low frequency modes form with their strongest vertical motion concentrated within the inner rings, in agreement with the above argument for clusters having large particle numbers.

TABLE I. Vertical modes for clusters with $N \leq 20$. The leftmost column lists the particle number $N$ while the top row gives the mode number, $l$. The first number for each entry lists $m$ and the second entry lists $n$.

|   | 4 | 5 | 6 | 7 | 8 | 9 | 10 | 11 | 12 | 13 | 14 | 15 | 16 | 17 | 18 | 19 | 20 |
|---|---|---|---|---|---|---|----|----|----|----|----|----|----|----|----|----|----|
| 4 | 2,1 | | | | | | | | | | | | | | | | |
| 5 | 2,1 | 0,1 | | | | | | | | | | | | | | | |
| 6 | 2,1 | 2,1 | 0,1 | | | | | | | | | | | | | | |
| 7  | 2,1 | 2,1 | 3,1 | 0,1 |     |     |     |     |     |     |     |     |     |     |     |     |     |
| 8  | 2,1 | 2,1 | 0,1 | 3,1 | 3,1 |     |     |     |     |     |     |     |     |     |     |     |     |
| 9  | 2,1 | 2,1 | 3,1 | 3,1 | 0,1 | 1,2 |     |     |     |     |     |     |     |     |     |     |     |
| 10 | 2,1 | 2,1 | 3,1 | 3,1 | 0,1 | 1,2 | 1,2 |     |     |     |     |     |     |     |     |     |     |
| 11 | 2,1 | 2,1 | 3,1 | 3,1 | 4,1 | 0,1 | 1,2 | 1,2 |     |     |     |     |     |     |     |     |     |
| 12 | 2,1 | 2,1 | 3,1 | 3,1 | 4,1 | 0,1 | 1,2 | 1,2 | 2,2 |     |     |     |     |     |     |     |     |
| 13 | 2,1 | 2,1 | 3,1 | 3,1 | 0,1 | 4,1 | 4,1 | 1,2 | 1,2 | 2,2 |     |     |     |     |     |     |     |
| 14 | 2,1 | 2,1 | 3,1 | 3,1 | 4,1 | 4,1 | 0,1 | 1,2 | 1,2 | 2,2 | --- |     |     |     |     |     |     |
| 15 | 2,1 | 2,1 | 3,1 | 3,1 | 0,1 | 4,1 | 4,1 | 5,1 | 1,2 | 1,2 | 2,2 | 2,2 |     |     |     |     |     |
| 16 | 2,1 | 2,1 | 3,1 | 3,1 | 0,1 | 4,1 | 4,1 | 5,1 | 1,2 | 1,2 | 2,2 | 2,2 | 0,2 |     |     |     |     |
| 17 | 2,1 | 2,1 | 3,1 | 0,1 | 3,1 | 4,1 | 4,1 | 5,1 | 5,1 | 1,2 | 1,2 | 2,2 | 2,2 | 0,2 |     |     |     |
| 18 | 2,1 | 2,1 | 3,1 | 3,1 | 0,1 | 4,1 | 4,1 | 5,1 | 5,1 | 1,2 | 1,2 | 2,2 | 2,2 | 3,2 | 0,2 |     |     |
| 19 | 2,1 | 2,1 | 3,1 | 0,1 | 3,1 | 4,1 | 4,1 | --- | --- | 6,1 | 1,2 | 1,2 | 2,2 | 2,2 | 3,2 | 0,2 |     |
| 20 | 2,1 | 2,1 | 0,1 | 3,1 | 3,1 | 4,1 | 4,1 | 5,1 | 5,1 | 6,1 | 1,2 | 1,2 | 0,2 | 2,2 | 2,2 | 3,2 | 3,2 |

## III CONCLUSIONS

In this work, the vertical oscillation modes as well as the horizontal oscillation modes for thermally excited 2D dust coulomb clusters were investigated. Both a box_tree simulation and an analytical method were employed to provide data for clusters with particle numbers between $N = 3$ and $N = 150$. The horizontal mode spectra were shown to agree with previously published results while the vertical mode spectra obtained from the box_tree simulation were compared with an analytical method and shown to agree with one another. The maximum frequency ($\omega_{z0}$) for the vertical mode oscillations was shown to be the vertical oscillation frequency for the system acting as a solid plane with the mode frequency decreasing as the mode number $l$ decreases. For clusters with small particle numbers ($N < 47$), all vertical modes show higher frequencies than corresponding horizontal modes. For larger clusters ($N \geq 47$), the vertical and horizontal modes are co-mingled.

Additionally, the resulting oscillation patterns for each of the vertical modes have been investigated. It was found that three highest frequency modes exist for all clusters examined ( $3 < N < 150$ ). The highest frequency mode is the mode corresponding to a vertical oscillation of the entire system of particles as a solid plane. This frequency is always equal to $\omega_{z0}$ and is independent of particle number. The modes with the second and third highest frequencies are quasi-degenerate and correspond to rotational oscillations of the system (again as a solid plane) around two different horizontal axes.

For clusters with large numbers of particles (N = 150, for example), the highest frequency modes after the three modes described above show oscillation patterns similar in shape to Bessel-Fourier functions with various indices of *m* and *n*. For specified values of *n*, modes with higher *m* values (except for $m = 0$) have lower frequencies and for specified values of *m*, modes with higher *n* values have lower frequencies. For every mode with $m \neq 0$, there is a second quasi-degenerate mode having the same *m* and *n* values with frequencies that are close but slightly different. This break in degeneracy is caused by the anisotropic nature of the cluster. At the other limit for the lowest frequency modes, the strongest particle motion within the cluster is concentrated within the first few inner rings with the outermost rings remaining almost motionless. In contrast, the horizontal modes show the strongest particle motion concentrated within the inner rings at their highest frequencies.[11]

For clusters with small numbers of particles ( $N \leq 20$ ), oscillation magnitudes for all modes appear in the shape of Bessel-Fourier functions. The mode order is almost the same as for large clusters except for a few specific cases which can be explained. As the particle number increases ( $N \geq 21$ ), additional low frequency modes appear showing strongest vertical motion concentrated within the inner few rings, including special modes where a peak-valley pair is concentrated at the center.

Figure 1 (Color online): Initial cluster structure for N = 6 (a), 46 (b) and 150 (c). Mode spectra for all normal modes (d-f) and vertical modes only (g-i) for N = 6 (d, g), 46 (e, h) and 150 (f, i).

Figure 2: Oscillation patterns for the three highest frequency vertical modes for clusters with $N = 3$ and $N = 150$.

Figure 3: Oscillation patterns for vertical modes with frequencies $\omega_{z0} \geq \omega \geq 14.32 s^{-1}$ ($N = 150$).

Figure 4: Oscillation patterns for vertical modes with frequencies $14.27 \geq \omega \geq 13.84 s^{-1}$ ($N = 150$).

Figure 5: Bessel functions for $m = 0$, $m = 1$, $m = 2$, and $m = 3$.

Figure 6: Oscillation patterns for the vertical modes with the lowest frequencies ($N = 150$).

Figure 7: Oscillation patterns for the three modes which can not be explained by the Bessel-Fourier function ($N \leq 20$).

Figure 8: Oscillation pattern for the lowest frequency vertical mode ($N = 21$).